\documentclass[twocolumn,aps,prl,showpacs]{revtex4}
\usepackage{graphicx}
\newcommand{\be}{\begin{equation}}
\newcommand{\ee}{\end{equation}}
\newcommand{\ben}{\begin{eqnarray}}
\newcommand{\een}{\end{eqnarray}}
\newcommand{\bb}{\bibitem}

\newcommand{\wt}{\widetilde}
\begin{document}
\title{Deformed Defects} 
\author{D. Bazeia$^a$, L. Losano$^a$ and J.M.C. Malbouisson$^b$} 
\affiliation{$^a$Departamento de F\'\i sica, Universidade Federal da
Para\'\i ba, 58051-970, Jo\~ao Pessoa, PB, Brazil\\
$^b$Instituto de F\'\i sica, Universidade Federal da Bahia, 
40210-340 Salvador, BA, Brazil} 
\date{\today} 

\begin{abstract}
We introduce a method to obtain deformed defects starting from a given
scalar field theory which possesses defect solutions. The procedure
allows the construction of infinitely many new theories that support defect
solutions, analytically expressed in terms of the defects of the original
theory. The method is general, valid for both topological and
non-topological defects, and we show how it extends to quantum mechanics,
and how it works when the scalar field couples to fermions. We illustrate
the general procedure with several examples, which support kink-like
or lump-like defects.
\end{abstract}


\maketitle

Defects play important role in modern developments of several
branches of physics. They may have topological or non-topological profile,
and in Field Theory the topological defects usually appear in models that
support spontaneous symmetry breaking, with the best known examples being
kinks and domain walls, vortices and strings, and monopoles \cite{hep1}.
Domain walls, for example, are used to describe phenomena having rather
distinct energy scales, as in high energy physics \cite{hep1,hep2}
and in condensed matter \cite{cm}.

The defects that we investigate in this letter are topological or kink-like
defects, and nontopological or lump-like defects. They appear in models
involving a single real scalar field, and are characterized by
their amplitude and width, the width being related to the region in space
where the defect solution appreciably deviates from vacuum states of the
system. Interesting models that support kink-like defects involve polynomial
potentials like the $\phi^4$ model, periodic potentials like the
sine-Gordon model, and even the vacuumless potential recently considered
in \cite{cvi,ba}. We shall investigate defects by examining their solutions
and the corresponding energy densities, to provide quantitative profile for
both topological and nontopological defects.

We introduce a general procedure to create deformed defects,
starting from a known solvable model in one spatial dimension.
We start with topological defects, and we show below that
the proposed scheme generates, for each given model having topological
solutions, infinitely many new solvable models possessing deformed
topological defects. We examine stability of kink-like defects,
to extend the procedure to quantum mechanics. We also investigate
lump-like defects, to generalize the procedure to both topological and
non-topological defects. Finally, we couple the scalar field to fermions,
to show how the procedure works for the Yukawa coupling. 

The interest in kink-like defects is directly related to the role of symmetry
restoration in cosmology \cite{hep1,hep2} and in condensed matter \cite{cm}.
Also, they are particularly important in other scenarios,
where they may induce interesting effects. A significant example
concerns the behavior of fermions in the background of kink-like
structures \cite{jre}. The main point here is that symmetry breaking
induces an effective mass term for fermions. In the background of the
kink-like structure the fermionic mass varies from negative to positive
values, and this fractionalizes the fermion number \cite{jre}. The
topological behavior of the kink-like defect is central to fermion number
fractionalization \cite{jre,gwi}. In the language of condensed matter,
spontaneous symmetry breaking may be interpreted as the opening of
a gap, and may be of good use in several situations -- see for
instance \cite{ssh,ols,rdj} and references therein for applications.
Another possibility concerns the role of kink-like defects as seeds for
the formation of non-topological structures \cite{c1,c2}. This line of
investigation has been implemented in the case the discrete symmetry is
changed to an approximate symmetry \cite{c3,c4}, and also when the symmetry
is biased to make domains of distinct but degenerate vacua spring
unequally \cite{c5}.

In our procedure to create deformed defects, we deform the system in a way
such that one increases or decreases the amplitude and width of the defect,
without changing the corresponding topological behavior. Within the
condensed matter context, one provides a way to increase or decrease
the mass gap for fermions, introducing an important mechanism to tune the
gap for practical purpose.

The interest in lump-like defects renews with the expressive number of recent
investigations on issues related to tachyons in String Theory,
since there are scenarios where branes may be seen as lump-like defects which
engender tachyonic excitations \cite{t1,t2,t3,t4,t5,t6,t7,t8,t9,t10}.

We consider a single real scalar field. The equation of motion for static
solutions $\phi=\phi(x)$ is given by
\be\label{em2}
\frac{d^2\phi}{dx^2}=V^{\prime}(\phi)
\ee
Here $V=V(\phi)$ is the potential, and the prime stands for derivative
with respect to the argument. We search for field configurations which
``start'' in a given minimum ${\bar\phi}$ of $V(\phi)$, with zero
``velocity'', that is, which obey the boundary conditions:
$\lim_{x\to-\infty}\phi(x){\to}{\bar\phi}$ and
$\lim_{x\to-\infty}d\phi/dx{\to}0$. Thus, we use
the equation of motion to get
\be\label{em1}
\frac{d\phi}{dx}=\pm\,\sqrt{2V(\phi)}
\ee
The energy associated with these solutions
are equally shared between gradient and potential portions
\ben
E&=&2E_g=\int_{-\infty }^{\infty }dx\,
\left(\frac{d\phi}{dx}\right)^2
\\
&=&2E_p=2\int_{-\infty }^{\infty}dx\,V(\phi)
\een

We now deal with topological or kink-like defects.
In this case we consider
\be\label{poteng1}
{V}=\frac{1}{2}\left[ W^\prime(\phi)\right]^{2}
\ee
where $W(\phi)$ is a smooth function of the field $\phi$. We assume that
there exist $v_{i},\,i=1,...,n$ such that $W^{\prime}(v_i)=0$.
These singular points of $W(\phi)$ are absolute minima of the potential.
In such a large class of models the equation of motion becomes
$d^{2}\phi/dx^2=W^{\prime}(\phi)W^{\prime\prime}(\phi)$.
The energy associated to $\phi(x)$ can be minimized to
\be
E^{\pm}_{BPS}=\pm \int^{\infty}_{-\infty} dx \,W^{\prime}(\phi)
\frac{d\phi}{dx}
\ee
if the field configuration obeys
\be\label{B}
\frac{d\phi_{\pm} }{dx}=\pm W^{\prime }(\phi_{\pm})
\ee
Their solutions are named BPS states \cite{ps,b}. As we know, for kink-like
defects the equation of motion exactly factorizes \cite{bms} into the two
first-order equations (\ref{B}). Thus, we can
introduce the topological current 
\be
J^{\alpha }=\epsilon ^{\alpha \beta }\partial _{\beta }W(\phi )\
\label{topcurr1}
\ee
which makes the topological charge equal to the energy of the topological
solution.

Let us now consider a well-defined bijective function $f=f(\phi)$ with
non-vanishing derivative. This function allows introducing a new theory,
defined by the $f$-deformed potential
\be
{\wt V}(\phi)=\frac{V[f(\phi)]}
{\left[f^{\prime}(\phi)\right]^2}=\frac12
\left(\frac{W^\prime[f(\phi)]}{f^{\prime}(\phi)}\right)^2
\label{dpotg}
\ee
In this case ${\wt v}_i=f^{-1}(v_i),\,i=1,2,...,n$ are minima,
and the new theory possesses topological defects which are
obtained from the solutions ${\phi}_{\pm}(x)$ of the previous theory
through the relation
\be
{\wt\phi}_{\pm }(x)=f^{-1}[{\phi}_{\pm}(x)]
\label{dsolg}
\ee
To prove this statement we notice that the first-order equations
of the new theory are
\be
\frac{d\phi}{dx}=\pm {\wt W}^{\prime}(\phi)=
\pm \frac{W^{\prime}(f(\phi))}
{f^{\prime }(\phi )}
\label{x3}
\ee
Thus, the solutions satisfy
$f[{\wt\phi}_{\pm}(x)]=\phi_{\pm}(x)$, or better
${\wt\phi}_{\pm}(x))=f^{-1}[\phi_{\pm}(x)]$, as written in
Eq.~(\ref{dsolg}).

We notice that the deformed defects
${\wt\phi}_{\pm}(x)$ connect minima corresponding to
those interpolated by the solutions ${\phi}_{\pm }(x)$ of the
original potential. The energy of the deformed defects depends on the
deformation one introduces. It can be written as
\ben
{\wt E}_{BPS}&=&\int_{-\infty }^{\infty}dx 
\left(\frac{d{\tilde \phi}}{dx}\right)^{2}\nonumber
\\
&=&\int_{-\infty }^{\infty }dx
\left(\frac{df^{-1}}{d\phi}\right)^{2}
\left(\frac{d\phi }{dx}\right) ^{2}\
\label{deneg}
\een
We see that for the class of deforming functions $f(\phi)$ satisfying
$|f^{\prime}(\phi)|\ge1$ ($\le1$), the energy is decreased (increased) relative
to the undeformed defect. In particular, the deformation $f(\phi)=r\phi$
leads to trivial modifications of parameters of the potential,
decreasing $(|r|>1)$ or increasing $(|r|<1)$ the energy of the defect.

At this point, two important remarks are in order: firstly, by taking
$f^{-1}$ instead of $f$ one defines the inverse deformation, that is
the $f^{-1}$-deformation of ${\wt V}(\phi)$ recovers the potential
$V(\phi)$. Secondly, the $f$- (or the $f^{-1}$-) deformation can be
applied repeatedly leading to an infinitely countable number of solvable
problems for each known potential bearing topological solutions. In fact,
each pair $(V,f)$ defines a class of solvable problems related to each other
through repeated applications of the $f$- (or $f^{-1}$-) deformation
prescription.

We concentrate on investigating stability of defects. This leads us
to quantum mechanics, where the Schr\"odinger-like Hamiltonian has the form
\be
H=-\frac{d^2}{dx^2}+U(x)
\ee
Here the quantum mechanical potential $U(x)$ is given by
\be
U(x)=\frac{d^2V(\phi)}{d\phi^2}\biggl|_{\phi=\phi(x)}
\ee
where $\phi(x)$ is the defect solution under investigation. In the case of
kink-like defects the potential is written as
$V(\phi)=(1/2)\,[W^{\prime}(\phi)]^2$, and the Hamiltonian can be
factorized \cite{ihu,cks} as $H=S^{\dag}\,S$, where the first-order operator
$S$ has the form
\be
S=\frac{d}{dx}+u(x)
\ee
and $u(x)=d^2W/d\phi^2$, to be calculated at the kink-like solution
$\phi=\phi(x)$. We use this to obtain the (bosonic)
zero mode in the form
\be
\eta_0(x)\sim e^{-\int^x dy\; u(y)}
\ee

We now use $f(\phi)$ to deform the model. The modified Hamiltonian
can be written as ${\wt H}={\wt S}^{\dag}{\wt S}$, where ${\wt S}$ is now
given by ${\wt S}=d/dx+{\wt u}(x)$, with
\be
{\wt u}(x)=W^{\prime\prime}(f(\phi))-
\frac{W^{\prime}(f(\phi))}{f^{\prime}(\phi)}\,
\frac{f^{\prime\prime}(\phi)}{f^{\prime}(\phi)}
\ee
to be calculated at the kink-like solution $\phi(x)$. Thus, the deformed
bosonic zero mode is given by
\be
{\wt\eta}_0(x)\sim e^{-\int^x dy\; {\wt u}(y)}
\ee

Let us now consider some examples. Firstly,
we consider $f(\phi)=\sinh\phi$, in which case the
$f$-deformation is referred to as the $\sinh$ deformation. For this choice,
equations (\ref{dpotg}), (\ref{dsolg}) and (\ref{deneg})
are easily rewritten and one sees that the Bogomol'nyi bound is lowered
by the deformation. On the other hand, if one considers the inverse
deformation, taking $f(\phi)={\rm arcsinh}\,\phi$, the energy of the
deformed defects is greater than that of the original potential.
Specifically, let us discuss the $\phi^{4}$ theory, for which the
potential is given by  $V(\phi)=(1/2)(1-\phi^2)^2$
(we take the rescaled theory with dimensionless field and coordinates).
The kink-like topological defects for this model are
well-known: ${\phi}^{(0)}_{\pm}(x)=\pm\tanh x$ (using the translation
invariance, we fix $x_{0}=0$). They have energy $E^{(0)}_B=4/3$, distributed
around the origin with density $\varepsilon _{0}(x)={\rm sech}^{4}(x)$.
In quantum mechanics, the related problem is described by the modified
P\"osch-Teller potential $U(x)=4-6\,{\rm sech}^2(x)$, which supports the
normalized zero-mode $\eta_0(x)=\sqrt{3/4}\;{\rm sech}^2(x)$ (at zero energy)
and another bound state, with higher energy.

The sinh-deformed $\phi^{4}$ model has potential given by
\be
{\wt V}(\phi)=\frac12{\rm sech}^2\phi(1-\sinh^2\phi)^2
\label{potdefphi4}
\ee
for which the deformed defects connecting the minima at
$\pm{\rm arcsinh}(1)$ are 
\be
{\wt\phi}^{(1)}_{\pm}(x)=\pm{\rm arcsinh}[\tanh(x)]
\label{solhdef}
\ee
See Fig.~[1] for a plot of the topological defects.

The ${\wt W}$-function for this example is given by
\be
{\wt W}(\phi )= 4\arctan(e^{\phi})-\sinh\phi  
\label{Wphi4h}
\ee
The deformed defects (\ref{solhdef}) have energy $E_{B}=(\pi-2)$, 
which is slightly smaller than the energy of the defects of the
$\phi^{4}$ potential. The energy density of the deformed defects
is given by
\be
{\wt\varepsilon}_{1}(x)=\frac{{\rm sech}^4(x)}{1+\tanh^2(x)}
\label{enerdenh}
\ee
which is more concentrated around the origin than the related quantity,
in the $\phi^{4}$ case, as expected. See Fig.~[2] for a plot of the energy
density of the topological defects.

Consider now the $\phi^4$ potential deformed with 
$f(\phi)={\rm arcsinh}\,\phi$, that is, take the potential
\be
{\wt V}(\phi)=\frac12 
(1+\phi^2)\left( 1-{\rm arcsinh}^2\phi\right)^2\;.
\label{adefphi4}
\ee
This is the potential which, by performing the deformation with $\sinh$
as discussed above, leads to the undeformed $\phi^4$ model.
The BPS solutions, in this case connecting minima at $\pm \sinh(1)$, 
are given by
\be
{\wt\phi}^{(-1)}_{\pm}(x)=\pm {\sinh}[\tanh(x)]
\label{solahdef}
\ee
These are deformed defects; see Fig.~[1].

The corresponding $W$-function is given by
\ben
{\wt W}^{(-1)}\phi)&=&-\frac16\,{\rm arcsinh}^3\phi 
+\left(\frac34+\frac12\phi^2\right){\rm arcsinh}\phi\nonumber 
\\
&&+\frac14\,\phi\sqrt{1+\phi^2}\left(1-2{\rm arcsinh}^2\phi\right) 
\label{arcW}
\een
The energy of the deformed defects in Eq.~(\ref{solahdef}) is
${\wt E}_B = 1.641$, which is greater than that for
the $\phi^4$ model and has a broader distribution
\be
{\wt\varepsilon}_{-1}(x)=\cosh^2(\tanh x)\,{\rm sech}^4(x)
\label{denahdef}
\ee
which is depicted in Fig.~[2].

\begin{figure}[h]
\includegraphics[height=5.5cm,width=8.0cm]{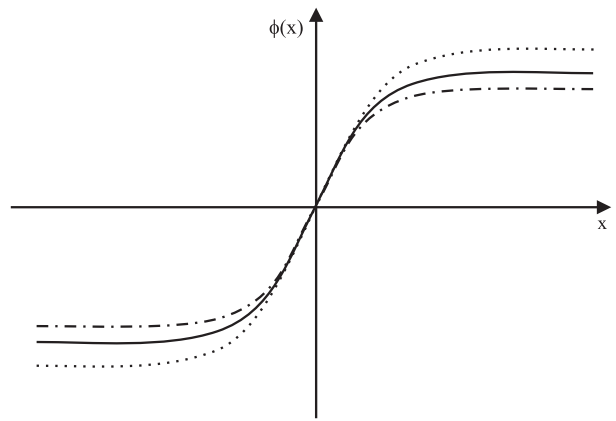}
\caption{Plot of the deformed defects. The thick line shows the kink of the
the $\phi^4$ model. The other lines show deformed kinks, the dashed-dotted
line representing the $\sinh$-deformation, and the dotted line the
${\rm arcsinh}$-deformation.} 
\end{figure}

We see that the $\sinh$-deformation diminishes the energy of the BPS
solutions narrowing its distribution, and the ${\rm arcsinh}$-deformation
operates in opposite direction, increasing the energy and spreading
its distribution. These deformations are smooth deformations, which lead
to potentials similar to the original potential. They map the interval
$(-\infty,\infty)$ into itself, and their derivatives $f^{\prime}(\phi)$
have no divergence at any finite $\phi$. They teach us how to deform a
given defect, changing its parameters in the two possible directions,
decreasing or increasing the amplitude and width of the original defect.
Since the amplitude and width of the defect are important to characterize
the defect, the proposed deformations are of direct interest to applications
involving kinks and walls in high energy physics and in condensed matter.

\begin{figure}
\includegraphics[height=5.0cm,width=8.0cm]{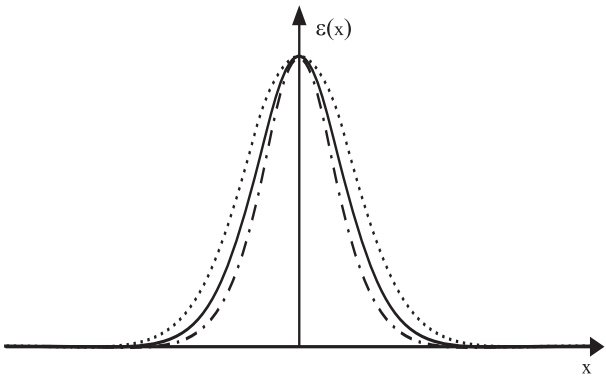}
\caption{Plot of the energy density of the deformed defects. The thick line
refers to the $\phi^4$ model. The other lines refer to the other
cases, as explained in the previous figure.} 
\end{figure}

The recent interest on tachyons \cite{t3,t4,t5,t6,t7,t8,t9,t10} has inspired us
to extend the above procedure to nontopological or lump-like defects.
We see that if $\phi(x)$ solves the equation
of motion (\ref{em2}), then ${\wt\phi}(x)=f^{-1}(\phi)$ solves the
equation of motion for the deformed model with potential
${\wt V}(\phi)=V[f(\phi)]/[f^{\prime}(\phi)]^2$.
This is always true, for solutions that obey the first-order Eqs.~(\ref{em1}),
with energy density equally shared between gradient and potential portions.

A model which supports nontopological or lump-like solutions is
\be
V_l(\phi)=\frac12\phi^2(1-\phi^2)
\ee
It has the solutions
\be
\phi^l_{\pm}(x)=\pm{\rm sech}(x)
\ee
In quantum mechanics,
the related problem has potential $U(x)=1-6\,{\rm sech^2(x)}$. This potential
has the same form of the the modified P\"osch-Teller potential [see the
comments just above Eq.~(\ref{potdefphi4})]. However, it plots differently,
shifting the values of $U(x)$ in a way such that the zero-mode is now
identified with the upper bound state, making the lower bound state negative,
signalling for tachyonic excitation.

We now consider deforming the lump-like solutions with $\sinh\phi$. We get
\be
{\wt V}_l(\phi)=\frac12\tanh^2\phi(1-\sinh^2\phi)
\ee
The equation of motion for $\phi=\phi(x)$ is
\be
\frac{d^2\phi}{dx^2}=\tanh\phi(1-\sinh^2\phi-2\tanh^2\phi)
\ee
It supports the deformed lump solutions
\be
{\wt\phi}^l_{\pm}=\pm{\rm arcsinh}[{\rm sech}(x)]
\ee
as we can verify straightforwardly. The deformation process may continue,
and may also be done in the reverse direction, using ${\rm arcsinh}\phi$.

Similar investigations apply to other potentials. For instance,
$V(\phi)=2\phi^2(1-\phi)$ supports the lump-like solution
$\phi^l(x)= {\rm sech}^2(x)$ -- see
Ref.~{\cite{t2}} for further details on the $\phi^3$ model. We deform the
lump-like solution with ${\rm arcsinh}\phi$. We get
\be
{\wt V}_l(\phi)=2(1+\phi^2){\rm arcsinh}^2\phi
(1-{\rm arcsinh}\phi)
\ee
The deformed lump-like defect is
\be
{\wt\phi}_l(x)=\sinh[{\rm sech}^2(x)]
\ee

We can make the model supersymmetric introducing appropriate Majorana spinors.
In this case, in general the Yukawa coupling is controlled by $Y(\phi)$, which
has the form
\be
Y(\phi)=\frac{d}{d\phi}\sqrt{2V(\phi)}
\ee
This leads to the usual coupling $Y(\phi)=W^{\prime\prime}(\phi)$ when the
potential is given by $V(\phi)=(1/2)\,[W^{\prime}(\phi)]^2$, which is the form
one uses to investigate kink-like structures. If one uses $f(\phi)$ to
change the model from $V(\phi)=(1/2)\,[W^{\prime}(\phi)]^2$ to
${\wt V(\phi)}=(1/2)\,[W^{\prime}(f(\phi))/f^{\prime}(\phi)]^2$, the Yukawa
coupling should also change from $Y(\phi)=W^{\prime\prime}(\phi)$
to
\be
{\wt Y}(\phi)=W^{\prime\prime}(f(\phi))-
\frac{W^{\prime}(f(\phi))}{f^{\prime}(\phi)}\,
\frac{f^{\,\prime\prime}(\phi)}{f^{\prime}(\phi)}
\ee

The importance of the deformation procedure that we have introduced enlarges
if one recognizes that it admits deformations which lead to very different
potentials, bearing no similarity to the original potential. Such deformations
are different, and may lead to further interesting situations. For instance,
we consider the function $f(\phi)=\tanh\phi$. It maps the interval
$(-\infty,\infty)$ into the limited interval $(-1,1)$, and this allows
introducing new effects, as we illustrate below.

We consider the potential
\be\label{ddpot}
V(\phi)=\frac12\,(1-\phi^2)^3
\ee
This potential is new. It is unbounded below, containing a maximum
at $\phi=0$ and two inflection points at $\pm1$. In this case the
modified potential becomes
\be\label{pot}
{\wt V}(\phi)=\frac12\,{\rm sech}^2\phi
\ee
which is the vacuumless potential considered in \cite{cvi,ba}. In
Ref.~{\cite{ba}} the vacuumless model was shown to support kink-like solutions
of the BPS type. This result indicates that the model (\ref{ddpot}) may also
support this kind of solutions. Indeed, it is astonishing to see that the
potential (\ref{ddpot}) supports the kink-like defects
\be\label{ddsol}
\phi(x)=\pm \frac{x}{\sqrt{1+x^2}}
\ee
which connect the two inflection points of the potential. These defects
are stable, and they can be seen as deformations of the defects
\be\label{sol}
\phi(x)=\pm{\rm arcsinh}(x)
\ee
which appear in the model defined by the potential of Eq.~(\ref{pot}). As
far as we know, this is the first example where kink-like defects connect
two inflection points. In the recent Ref.~{\cite{jva}} one has found another
model, somehow similar to the above one, but there the solution connects a
local minimum to a inflection point.

The solutions (\ref{ddsol}) are stable, and the Schr\"odinger-like
equation that appears in the investigation of stability is defined by
the Hamiltonian
\be
H=-\frac{d^2}{dx^2}+12\frac{x^2-1/4}{(x^2+1)^2}
\ee
The potential is a volcano-like potential, which supports the zero mode
and no other bound state. The (normalized) wave function of the zero mode
is $\eta_0(x)=2(2/3\pi)^{1/2}(x^2+1)^{-3/2}$. This should be contrasted
with the zero mode of the vacuumless potential, which is given by \cite{ba}:
${\wt\eta}_0(x)=(1/\pi)^{1/2}(x^2+1)^{-1/2}$. We notice that the two
zero modes localize very differently in space.

The present work is of direct interest to investigations concerning
systems described by two real scalar fields, as considered for instance
in Refs.~{\cite{2f,bb}}. Also, it may be of some use in more complex
situations, involving three or more scalar fields, in scenarios such
as the one where we deal with the entrapment of planar network of
defects \cite{bb01}, or with the presence of non-trivial solutions
representing orbits that connect vacuum states in the three-dimensional
configuration space \cite{blw}.

The deformation scheme that we have presented may also work in other
contexts, in particular in the case where one couples the scalar field
to gravity in higher dimensions. We have found interesting investigations
in Refs.~{\cite{g1,g2,g3}}, and we are now considering the possibility of
extending the deformation procedure to brane-world scenarios.

We would like to thank C.G. Almeida, F.A. Brito, and R. Menezes for
discussions, and CAPES, CNPq, PROCAD and PRONEX for partial support.


\end{document}